# Current-induced Nonequilibrium Phase Transition Accompanied by Giant Gap Reduction in Vanadium Dioxide

Akitoshi Nakano*, Masato Imaizumi, and Ichiro Terasaki

*Department of Physics, Nagoya University, Nagoya 464-8602, Japan.*

**Abstract**

We investigated nonlinear conduction in bulk single crystals of $VO_2$ with precise temperature control. Two distinct nonequilibrium phenomena were identified: a gradual reduction of the charge gap and a current-induced insulator–metal transition. The electric field required to drive the nonlinear conduction is two to three orders of magnitude smaller than that reported for $VO_2$ thin films or nanobeams, strongly indicating an intrinsic electronic origin rather than a temperature increase due to self heating. Notably, our results suggest that the application of a steady current to the frozen insulating state can induce a nonequilibrium steady-state metallic phase—effectively "melting" the electronic ice. This highlights a novel route to controlling electronic states via nonthermal, current-driven mechanisms.

**Main text**

Nonlinear conduction—where the current–voltage relationship deviates from Ohm's

law—is an intriguing transport phenomenon in solids. In semiconductors, where the single-electron approximation holds, nonlinear conduction is known to arise from mechanisms such as hot electron effects [1], the Gunn effect [2], impact ionization [3], and Zener tunneling [4], and is exploited in devices including oscillators, detectors, multipliers, and diodes. Typically, such phenomena occur under high electric fields on the order of $10^3$-$10^7$ V·cm$^{-1}$ at room temperature. In contrast, nonlinear conduction can also emerge in insulating systems where a charge gap forms due to many-body interactions. A well-known example is the sliding of charge density waves above a threshold electric field [5]. In recent years, current-induced phase transitions and anomalous transport behaviors under current flow have been reported in various correlated electron systems, including organic conductors [6–8], prototypical Mott insulators [9], and orbital-ordered Mott insulators [10–14]. Remarkably, these phenomena occur under very low electric fields—on the order of tens of V·cm$^{-1}$—suggesting a direct current (dc) voltage or current effect fundamentally distinct from conventional nonlinear conduction in semiconductors. Such current-controlled insulator–metal transitions not only enable novel functionalities but also offer a platform for extending the framework of equilibrium thermodynamics to the nonequilibrium regime. However, because nonlinear conduction is inherently accompanied by Joule heating [15–17], it remains debated whether these effects are intrinsic to the electronic system or simply a consequence of local temperature rise.

Vanadium dioxide ($VO_2$) is a canonical material for investigating nonlinear conduction, yet its underlying mechanism remains under debate. $VO_2$ undergoes a dramatic resistance increase—by 3 - 5 orders of magnitude—below 340 K [18], accompanied by the structural phase transition from the Rutile ($P4_2/mnm$) phase to the M1 ($P2_1/c$) phase. The formation of the molecular orbital state of vanadium dimers is one of the most representative characteristics of M1 phase [19]. Additionally, the high-temperature Rutile phase exhibits a structurally unstable metallic state [20–22], and a variety of metastable phases have been reported under slight perturbations such as uniaxial stress [23] or elemental substitution [24,25]. These observations suggest that multiple competing instabilities coexist in this chemically simple compound. Due to the potential for externally controlling its unique electronic states, the nonlinear conduction properties of $VO_2$ have attracted sustained attention. Several studies on thin films and nanobeams have reported voltage-induced conductance switching [26], current oscillations [27], and metallic phases retaining dimerized structures [28,29], supporting an interpretation based on nonthermal electronic effects. On the other hand, filamentary conduction paths visualized using optical microscopy or blackbody radiation mapping suggest that Joule heating may be dominant in some cases [30,31]. Thus, the nature of nonlinear conduction in $VO_2$ remains controversial. Importantly, nearly all prior studies have focused on thin films or nanobeams, which are subject to substrate and surface effects. To the best of our knowledge, there have been no systematic investigations of nonlinear conduction in bulk single crystals with accurate temperature evaluation to isolate intrinsic behavior.

In this study, we performed nonlinear conduction measurements on bulk single crystals of $VO_2$, carefully accounting for sample temperature. Two types of nonequilibrium responses were identified: (i) a continuous reduction in the charge gap, and (ii) a current-induced first-order insulator–metal transition. Notably, the electric field required to induce these effects in our bulk crystals is 2–3 orders of magnitude smaller than that reported in thin films or nanobeams, providing strong evidence for an intrinsic, nonthermal origin.

Single crystals of $VO_2$ were synthesized by melting $V_2O_5$ at 950 °C in a nitrogen gas flow (4 L/min) for three days, following the method reported in Ref. [32], by using platinum cruicibles and a tube furnace. The resulting crystals exhibited an anisotropic plate-like morphology [Fig. 1(a)], with the long axis corresponding to the *c*-axis of the rutile structure. The crystal quality was confirmed using synchrotron singele crystal X-ray diffraction. The composition determined from the analysis is consistent with the stoichiometric ratio to within 0.2%, indicating high compositional precision. A sharp insulator–metal transition near 340 K was observed in the temperature-dependent resistivity [Fig. 1 (b)], again suggesting that the chemical composition is very close to the stoichiometric ratio. The experimental setup for nonlinear conduction measurements is illustrated in Fig. 1 (c). To accurately monitor the sample temperature under high current densities, we employed the non-contact infrared thermometer (Japan Sensor, TMHX-CSE0500-0100E001) as developed by Okazaki et al. [11]. This device detects blackbody radiation in the infrared range ($\lambda$ = 6–12 μm), thereby enabling temperature measurements that are free from the added heat capacity and thermal contact resistance associated with conventional contact-type thermometers. In our setup, the $VO_2$ crystal was suspended using four gold wires (25 μm diameter) to thermally isolate it from the sample stage and suppress temperature gradients between the upper and lower surfaces. The sample temperature was controlled via a Peltier module in conjunction with a current generator (Matsusada Precision, P4K18-2). The infrared thermometer readings were calibrated against a platinum resistance thermometer attached to the Peltier stage. As shown in Fig. 1 (d), the agreement between the two measurements was within 3 K below 340 K. The resistivity along the long axis was measured using a standard four-probe DC technique under various applied currents supplied by a current source (Keithley 6221). To avoid thermal runaway typically associated with voltage-controlled nonlinear conduction, DC current was used as the control parameter. To minimize contact resistance—an extrinsic source of Joule heating and apparent nonlinearity—a thick gold layer (~1000 Å) was deposited on the contact regions of the crystal surface before bonding the gold wires using silver paste. Under the applied current conditions, the temperature rise at the center of the sample reached at most ~50 °C, which remains significantly below the insulator–metal transition temperature of $VO_2$.

Figure 2 (a) shows the temperature dependence of resistivity ($\rho$) measured under various current densities ($J$). The semiconducting behavior, characterized by negative temperature dependence (d$\rho$/d$T$ < 0), is gradually suppressed with increasing $J$. It should be emphasized that the plotted temperature corresponds to the actual surface temperature of the sample, determined by the infrared thermometer, rather than the stage temperature. At higher currents, data at low temperatures are unavailable due to significant self-heating, which prevented sufficient cooling. While previous studies have attributed electric-field-induced insulator–metal transitions in $VO_2$ to Joule heating, our results strongly suggest that nonlinear conduction of intrinsic, nonthermal origin occurs in bulk single crystals. In particular, we observed a sudden drop in resistivity at a certain temperature for $J$ exceeding ~3.0 A·cm$^{-2}$. Notably, this abrupt change in resistivity exhibits thermal hysteresis, as shown in the inset of Fig. 2(a), indicating that the phenomenon is due to a first-order phase transition rather than the creation of oxygen vacancies. The resulting $J$–$T$ phase diagram is summarized in Fig. 2 (b).

We now discuss the two distinct types of current-induced nonequilibrium phenomena observed in $VO_2$. The first is the continuous reduction of the activation gap in the low-current regime. Figure 3 (a) shows the Arrhenius plot of the $\rho$–$T$ data from Fig. 2 (a), limited to the data prior to the abrupt transition. These plots follow an ideal thermally activated behavior, with the slope—corresponding to the activation energy—clearly decreasing with increasing $J$. Using the conventional expression $\rho(T) = \rho_0 \cdot \exp(\Delta/2k_BT)$, where $\rho_0$, $\Delta$, and $k_B$ are the pre-exponential factor, activation energy, and Boltzmann constant, respectively, we extracted the gap energy as a function of current [Fig. 3 (b)]. The estimated gap at the lowest current ($J = 0.1$ A·cm$^{-2}$) is ~ 1 eV, in reasonable agreement with reported values (~0.7 eV) [33]. As $J$ increases, the gap gradually decreases to ~0.5 eV at $J \sim 2$ A·cm$^{-2}$, and then exhibits saturation-like behavior near 3–5 A·cm$^{-2}$. Although similar current-induced gap reductions have been observed in Ca$_2$RuO$_4$ [9] and organic conductors [8], the magnitude of the gap modulation in VO$_2$ is an order of magnitude larger. Remarkably, such a substantial reduction of several tenths of an eV occurs under electric fields as low as a few tens of V·cm$^{-1}$. Since gap modulation in other systems is linked to the collapse of orbital or charge order, our results suggest that orbital ordering and/or electronic correlations are likely involved in the insulator–metal transition of VO$_2$, consistent with prior theoretical studies [34] and X-ray absorption measurements [35]. These behaviors were consistently reproduced in three different crystals, supporting the conclusion that this phenomenon is an intrinsic nonequilibrium response of bulk VO$_2$.

The second nonequilibrium phenomenon observed is a current-induced first-order insulator–metal transition, which occurs when the activation gap becomes reduced to ~ 0.7 eV. Figure 4 (a) shows reconstructed $\rho$–$J$ curves at various temperatures based on the data in Fig. 2 (a). As the current increases, the resistivity decreases smoothly due to the continuous gap reduction. At the maximum applied current, $\rho$ drops to about 3% of its value at the lowest current at 310 K. The $E$–$J$ characteristics, shown in Fig. 4 (b), exhibit negative differential resistance (NDR) behavior emerging near $J$ = 2–3 A·cm$^{-2}$, which corresponds to the onset of the strong resistivity suppression. Notably, abrupt drops in the $E$–$J$ curves are clearly observed at each temperature for $J \approx$ 3–5 A·cm$^{-2}$, indicating that a current-induced first-order transition is taking place. Such features have not been previously reported in VO$_2$, despite extensive investigations of electric-field-induced insulator–metal transitions.

Importantly, the onset of the transition in the $\rho$–$J$ curves occurs at nearly the same resistivity value (~16 Ω·cm), regardless of temperature. This is also reflected in the $E$–$J$ plots, where the transition points align along a straight line [Fig. 4 (b)], implying that the threshold condition is governed by resistivity rather than electric field strength. In contrast, the threshold electric field ($E_{PT}$) gradually decreases with decreasing temperature, as shown in the inset of Fig. 4 (a). This behavior is in stark contrast to the sliding of charge density waves, where the critical electric field typically serves as the control parameter. These observations suggest that resistivity —a nonthermodynamic parameter—may act as an effective order parameter for the current-induced transition in VO$_2$, as previously proposed for V$_2$O$_3$ [36].

We now turn to the nature of the nonequilibrium state that emerges after the current-induced insulator–metal transition in bulk VO$_2$. This suggests that the application of a steady current to the frozen insulating state can induce a nonequilibrium steady-state metallic phase—effectively "melting" the electronic ice. Recent optical microscopy studies have revealed the formation of coexisting metallic and insulating domains under DC voltage bias in single crystals [37,38]. Interestingly, the observed domain boundaries are oriented obliquely with respect to the electric field, in contrast to the filamentary conduction paths reported in thin films. Our ongoing X-ray diffraction measurements further confirm that a monoclinic dimerized structure coexists with the rutile metallic phase above a certain current density, with a detectable volume fraction [39]. The emergence of metastable phases with slightly different structures may relieve internal stress in the single crystal, leading to complex, multidirectional domain patterns. In contrast, in VO$_2$ thin films and nanobeams, once a high electric field ($\sim 10^5$ V·cm$^{-1}$) induces a filament connecting the electrodes, the system undergoes a complete transition into the metallic state due to intense Joule heating [30], leaving no room for phase coexistence. To quantify this distinction, we estimate the characteristic correlation length $l$ of the nonlinear transition using the relation $l = \delta\Delta / (eE_0)$, where $\delta\Delta \sim 0.7$ eV is the gap reduction at the threshold current $J_{PT} = 4$ A·cm$^{-2}$, and $E_{PT} \sim 70$ V·cm$^{-1}$ at 310 K (sample #1). This yields $l \sim 100$ μm, consistent with the sub-millimeter domain sizes observed in Refs. [37,38].

We infer that this long correlation length is responsible for the distinct nonlinear conduction behavior in bulk crystals compared to thin films or nanobeams. Since typical film and nanobeam dimensions are smaller than this length scale, they require much larger electric fields to induce phase transitions. Furthermore, this correlation length is far longer than the electronic mean free path, implying that the electronic system couples strongly to lattice degrees of freedom. This scenario is reminiscent of electron–phonon–coupled transport reported in $Ca_2RuO_4$ [11], where lattice deformations and domain formation are easily disrupted by substrate-induced stress. While this self-organized domain structure is not identical to the intrinsic inhomogeneity observed in organic conductors [6], high-$T_c$ cuprates [40], or relaxor ferroelectrics [41], it shares characteristics with long-range correlated self-organization like "dissipative structure" under a nonequilibrium condition. At present, the microscopic mechanism underlying this current-induced self-organization remains unclear. Further theoretical and experimental investigations are needed to fully elucidate this unique nonequilibrium phenomenon.

In summary, we have investigated nonlinear conduction in bulk single crystals of $VO_2$ with precise evaluation of the sample temperature. The resistivity, measured via a four-probe method, was continuously suppressed with increasing applied current, reflecting a reduction in the charge gap. The activation energy decreased from ~ 1 eV at $J = 0.1$ A·cm$^{-2}$ to ~ 0.5 eV at $J = 2$ A·cm$^{-2}$. Furthermore, an abrupt drop in resistivity was observed at each temperature when the resistivity reached a threshold value of ~ 16 Ω·cm, indicating a current-induced first-order insulator–metal transition. The estimated correlation length for the transition is ~ 100 μm, suggesting a strong coupling between electronic and lattice degrees of freedom via electron–phonon interaction. This long length scale likely accounts for the striking contrast in nonlinear conduction behavior between bulk single crystals and thin films or nanobeams. Our findings highlight the existence of an intrinsic, nonthermal mechanism for nonlinear conduction and current-induced phase transitions in $VO_2$, and reveal the importance of spatial correlations in the nonequilibrium dynamics of strongly correlated electron systems.

**Acknowledgement**

This work was supported by a Grant-in-Aid for Science Research (B)(proposal No. 22H01166 & 23K22437) from the Japan Society for the Promotion of Science (JSPS).

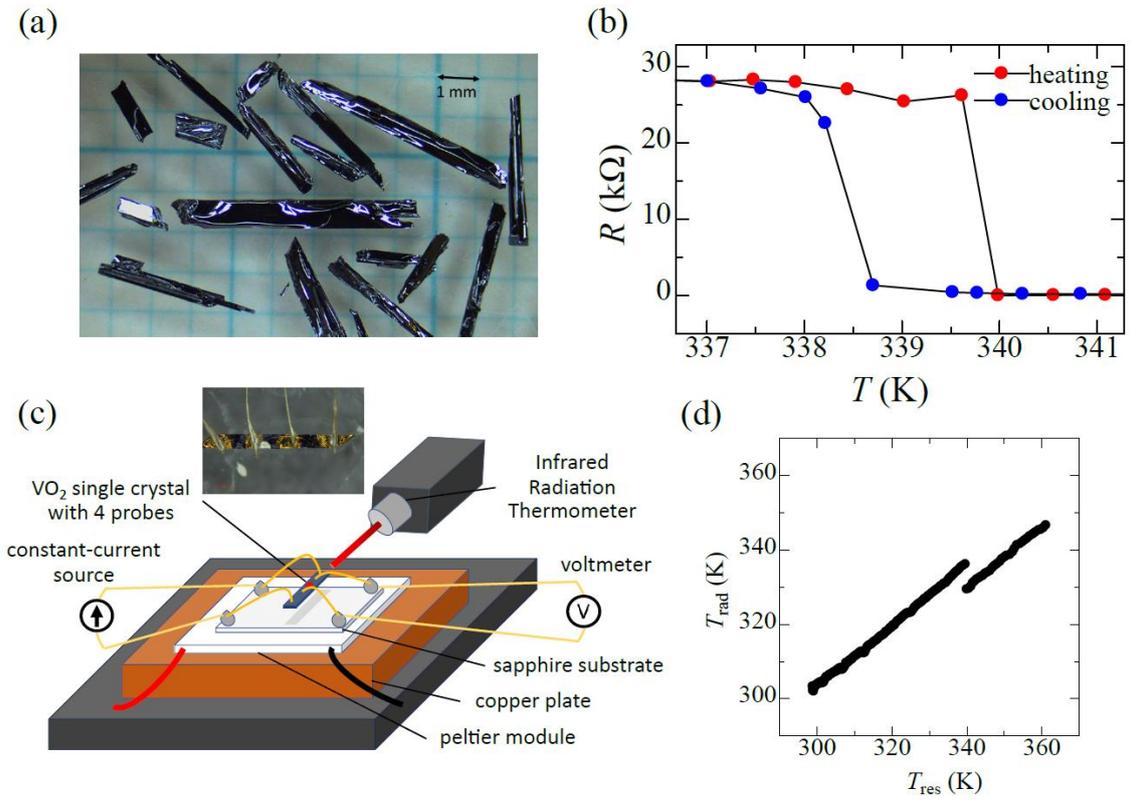

Figure 1 (a) Single crystals of VO$_2$. (b) Confirmation of the metal-insulator transition at $T \sim 340$ K. (c) Measurement apparatus of non-linear conductivity. (d) Comparison of the temperatures determined by Pt resistor ($T_{res}$) and the infrared radiation thermometer ($T_{rad}$). $T_{rad}$ jumps at $T \sim 340$ K, indicating the change in reflectivity associated with the metal-insulator transition.

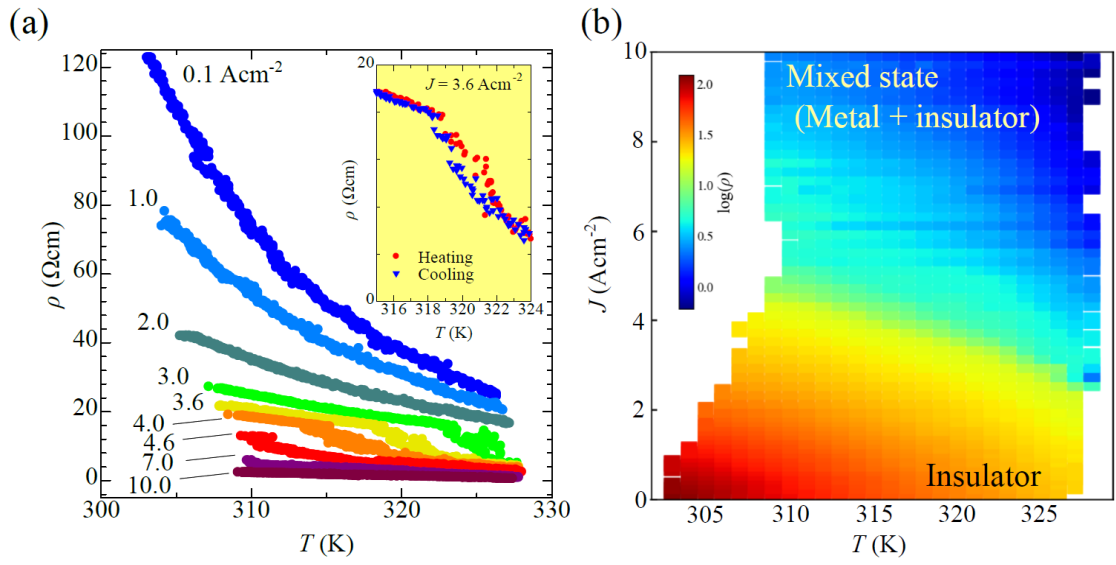

Figure 2 (a) Temperature dependence of resistivity ($\rho$) under various applied current density ($J$). The inset shows thermal hysteresis of the first-order phase transition at $J = 3.6$ A·cm$^{-2}$. (b) $J$ - $T$ phase diagram constructed by using $\rho$ - $T$ data in (a).

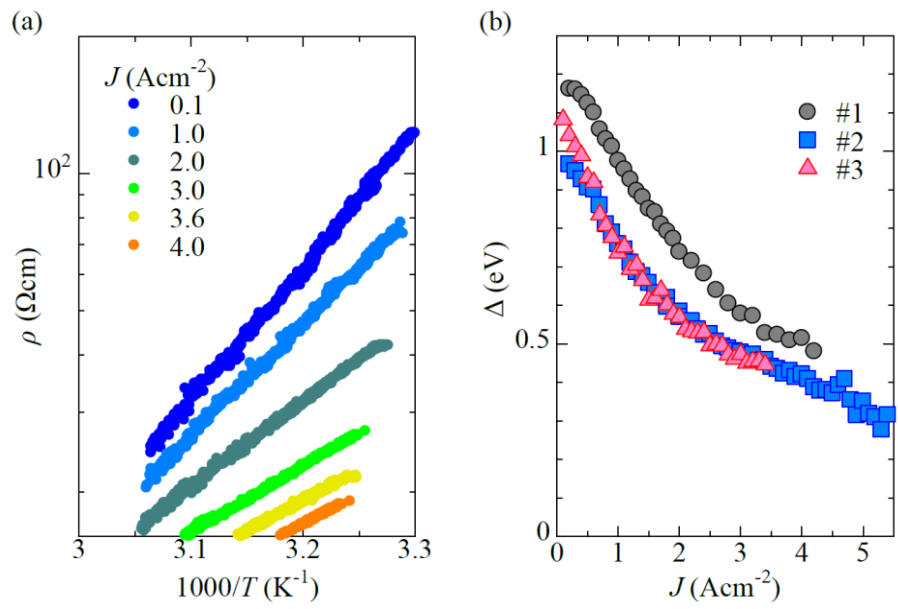

Figure 3 (a) Arrhenius plot of the temperature dependent $\rho$ under various $J$. (b) $J$-dependent activation energy ($\Delta$). #1~#3 denote the sample numbers collected from the same batch.

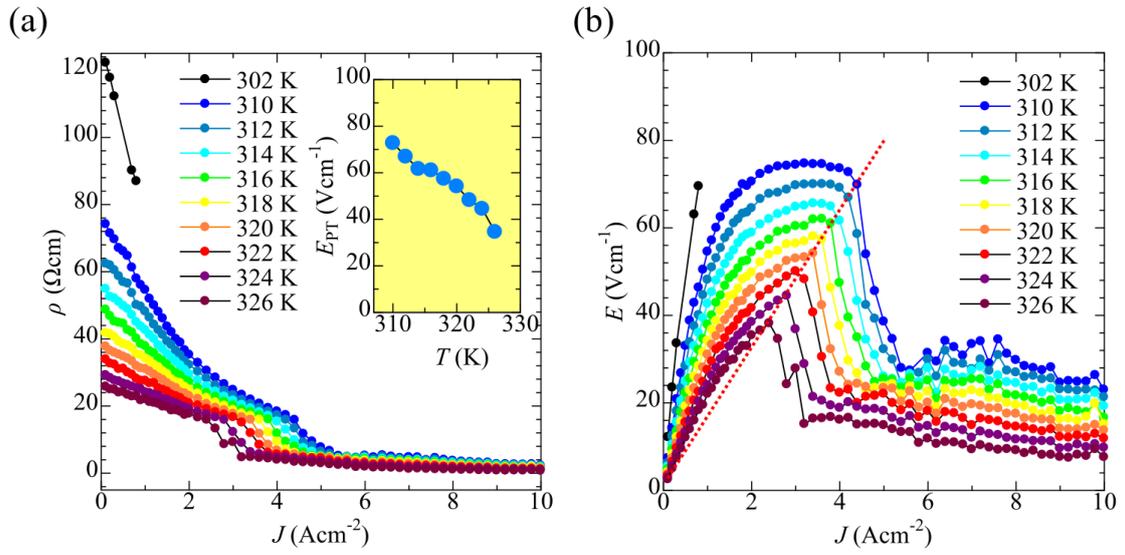

Figure 4 (a) *J*-dependence of $\rho$ at various temperatures. The inset shows the onset electric-fields (*E*) and $\rho$ of the current-induced nonequilibrium phase transition. (b) *J*-dependence of *E* at various temperatures. The red line indicates $E = \rho J$ relatiship, where $\rho = 16$ $\Omega$cm.